Research Article

# Multiple Bipolar Fuzzy Measures: An Application to Community Detection Problems for Networks with Additional Information

Inmaculada Gutiérrez[1,*], Daniel Gómez[1,2], Javier Castro[1,2], Rosa Espínola[1,2]

[1]*Faculty of Statistics, Complutense University Puerta de Hierro, s/n, Madrid, Spain*
[2]*Instituto de Evaluación Sanitaria, Complutense University, Madrid, Spain*

**ABSTRACT**

In this paper we introduce the concept of multiple bipolar fuzzy measures as a generalization of a bipolar fuzzy measure. We also propose a new definition of a group, which is based on the multidimensional bipolar fuzzy relations of its elements. Taking into account this information, we provide a novel procedure (based on the well-known Louvain algorithm) to deal with community detection problems. This new method considers the multidimensional bipolar information provided by multiple bipolar fuzzy measures, as well as the information provided by a graph. We also give some detailed computational tests, obtained from the application of this algorithm in several benchmark models.

## 1. INTRODUCTION

In general, relations, communication or even feelings among humans, have two aspects, the positive and the negative. The first one expresses possible or permitted evidences which provide satisfaction. On the contrary, the negative aspect represents impossible relations or situations which are unacceptable or not permitted, and provide dissatisfaction [1].

When some message arrives to human consciousness, the mind automatically assigns to it negative and positive evidences. Being able to deal in this way with the information, while simultaneously considering positive and negative evidences, humans are capable of representing and handling with complex situations and scenarios connected to sensations, ambivalence, conflicts, or interests [2]. This peculiar manner of managing information has its basis on the idea of bipolarity, which was originally a psychological concept [3].

In this work we take into account these considerations, adding them in the field of graphs or networks, which are one of the most important fields in which clustering methods are applied. Given a finite set of objects, graphs are used to model and visualize the relations or connections among its elements. The complexity of these problems grows almost exponentially with the consideration of models requiring hundreds or thousands of individuals and edges, so it becomes very more important to build robust and effective systems to choose the most appropriate algorithms for defining blocks.

Communities can be understood as groups of related individuals in social networks [4], biochemical pathways in metabolic networks [5], or even sets of Web pages dealing with the same topic [6]. Apart from their wide range of applications, communities can be used to visualize how the society works. A community structure can also be used to find groups of elements depending on their relations. As far as possible, we assume that a group should be composed of elements between which there is some affinity, trying to keep separate those elements between which there is some discrepancy.

In this paper, we follow the philosophy introduced in [7], which consists in considering some bipolar additional information independent of the topology of the graph for community searching. Doubtlessly, the connections among the individuals, modeled by the edges of the graph, should be the main factor in the search of a cluster structure. However, it would be reasonable to consider some additional information apart from that provided by the structure of the graph, in order to extend the notion of community. Moreover, this double perception of the information may be given by several sources. Here lies the importance of working with multiple bipolar fuzzy measures.

[*]*Corresponding author. Email: inmaguti@ucm.es*

Let us show an example with 8 individuals whose connections can be seen in Figure 1. The set of edges is $E = \{\{1,2\},\{1,4\},\{2,3\},\{3,4\},\{4,6\},\{5,6\},\{5,8\},\{6,7\},\{7,8\}\}$. Considering the information provided by the structure of the graph, any community detection problem will easily find two modules or clusters, $C_1 = \{1,2,3,4\}$ and $C_2 = \{5,6,7,8\}$. However, let us imagine we have two information sources: one is about the personal relationships between these individuals, so that there are some pairs of individuals who are enemies (1 and 4; 6 and 8), and there are some pairs of individuals who are old friends (1 and 2; 3 and 4; 5 and 6); the other one is about working relations, so we know the relation between some individuals who have opposite work interests thus they have bad relations (1 and 3; 2 and 4; 5 and 7; 6 and 7), whereas there is a pair of individuals who are associate with a tight relation (7 and 8). Then, if we reformulate the notion of a group considering all this additional information about the different relations of affinity/discrepancy among the individuals, it seems logical that there should be four groups, $C_1 = \{1,2\}$, $C_2 = \{3,4\}$, $C_3 = \{5,6\}$, $C_4 = \{7,8\}$. Let us remark that, although in this case the clustering is based on symmetrical relationships (enemies and old friends or extreme working relations), we could also apply this method when relations are nonsymmetrical. To deal with that, we have to measure how each relation affects to the concept of group and represent it mathematically.

According to our best knowledge, this way to manage the additional information with a double perception in a multiple case, is an approach which has never been deeply studied. Recent studies have dealt with the problem of incorporating some additional information when finding a cluster structure in a graph [8–11]. A preliminary work focused on improving the community detection problem by incorporating the use of fuzzy measures, was published in [12]. This proposal is based on the consideration of fuzzy measures defining affinity relations, so that we know which nodes should be in the same cluster.

Later, in [7], it was proposed the use of a bipolar fuzzy measure to find communities in a network in a more realistic way. Hence, considering this bipolar fuzzy measure (in a unidimensional case), not only the positive information, but also the negative one is analyzed to define clustering structures.

Now we come up with the use of multiple bipolar fuzzy measures, which allow us to approach complex scenarios that cannot be modeled with those tools exploited in previous works. In this background, we know not only which elements should be in the same group because of their affinity, but also which ones should be separated due to their discrepancy. This approach is somewhat similar to intuitionistic fuzzy cognitive maps (FCMs) [13,14] (in a multidimensional scale), that describe the connections between nodes by membership/nonmembership values.

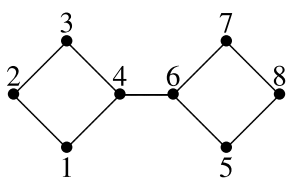

**Figure 1** | Example. Graph with 8 nodes.

Then, in this work we propose an extension of the Louvain algorithm [15] which can consider two types of input information. One is about the direct connections between the nodes, given by the edges of the graph. It is used to find "possible" clusters. The other, which could be also the edges of the graph or even an independent information, is used to search the maximum of modularity. This new approach, called *Duo Louvain*, is formalized in Algorithm 2. Then, we propose a particular application of *Duo Louvain* Algorithm, named *Multiple Bipolar Duo Louvain* Algorithm, for searching community structures in extended multiple bipolar fuzzy graphs. It is formalized in Algorithm 3. There, in a multidimensional case, we develop a way to manage bipolar information which shows us which nodes should be separated, and which ones should remain together.

We test the "goodness" of our proposal with the use of different benchmark models [16], in combination with the calculation of the *Normalized Mutual Information* (*NMI*) [17]. Having a standard or "gold" group structure, the objective is to quantify the ability of our algorithm to recover the embedded partition. Results allow us to guarantee the good performance of our algorithm when working with extended multiple bipolar fuzzy graphs.

The remainder of the paper is organized as follows: In Section 2 we describe the concepts we will use along the paper, including the definition of extended bipolar fuzzy graphs, the bipolar weighted multi-graph associated with a bipolar fuzzy measure, and some definitions of aggregation operators (AOs). In Section 3, we introduce some key concepts: the extended multiple bipolar fuzzy graph and the bipolar weighted multi-graph associated with multiple bipolar fuzzy measures. In this section we also describe the aggregation process proposed to deal with this complex information. Section 4 is devoted to the Louvain algorithm, including a detailed explanation about its performance in Subsection 4.1, and a modification of it to deal with community detection problems based on multiple bipolar fuzzy measures, in Subsection 4.2. We test the performance of our proposal with some computational results in Section 5. To finish the paper, we provide some conclusions and final remarks in Section 6.

## 2. PRELIMINARIES

### 2.1. AOs and Other Functions

AOs [18] are one of the hottest disciplines in information sciences. AO appear in a natural way when the soft information has to be aggregated. At the beginning, AO were defined to aggregate values from membership functions associated with fuzzy sets [19]. AOs, also known as aggregation functions, are a key concept for the development of this paper.

**Definition 1. Aggregation Function** [18]. The function $A : [0,1]^n \to [0,1]$ is said to be an $n$-ary aggregation function if the following conditions hold:

1. $A$ is increasing in each argument: for each $i \in \{1,\ldots,n\}$, if $x_i \leqslant y$, then $A(x_1,\ldots,x_n) \leqslant A(x_1,\ldots,x_{i-1},y,x_{i+1},\ldots,x_n)$.

2. $A$ satisfies the boundary conditions: $A(0,\ldots,0) = 0$ and $A(1,\ldots,1) = 1$.

Let us note that aggregation functions are usually classified into four different classes, by means of their point-wise comparison to the maximum and minimum operators. These classes are as follows:

- **Disjunctive aggregation function**: an aggregation function $A_n : [0,1]^n \to [0,1]$ is said to be disjunctive if and only if, for every $x = (x_1, \ldots, x_n) \in [0,1]^n$, then $A_n(x) \geqslant \max\{x_i, \ 1 \leqslant i \leqslant n\}$.
- **Conjunctive aggregation function**: an aggregation function $A_n : [0,1]^n \to [0,1]$ is said to be conjunctive if and only if, for every $x = (x_1, \ldots, x_n) \in [0,1]^n$, then $A_n(x) \leqslant \min\{x_i, \ 1 \leqslant i \leqslant n\}$
- **Averaging aggregation function**: an aggregation function $A_n : [0,1]^n \to [0,1]$ is an averaging aggregation function if and only if, for every $x = (x_1, \ldots, x_n) \in [0,1]^n$, then $\min\{x_i, \ 1 \leqslant i \leqslant n\} \leqslant A_n(x) \leqslant \max\{x_i, \ 1 \leqslant i \leqslant n\}$.
- **Mixed aggregation function**: an aggregation function $A_n : [0,1]^n \to [0,1]$ that does not belong to any of the three previous classes is said to be a mixed aggregation function.

Finally, let us define the negation operator that will be used later in this paper.

**Definition 2. Negation Operator** [20]. Given a partial ordered set $(X, <)$, a negation function $N$ is defined as a function from $X$ to $X$ such that the following two conditions hold:

(N1) If $a < b$ then $N(b) < N(a)$, for every $a, b \in X$.

(N2) $N(N(a)) = a$, for every $a \in X$.

## 2.2. Fuzzy Graphs, Bipolar Fuzzy Graphs, and Weighted Graphs Associated with Fuzzy Measures

Let the pair $G = (V, E)$ be a graph or network, where $V = \{1, \ldots, n\}$ is the set of nodes, and $E = \{\{i,j\} \mid i,j \in V\}$ is the set of edges or arcs. We assume that two nodes $i, j$ are related in $G$ if $\exists \{i,j\} \in E$. Let $A$ be the adjacency matrix of $G$, showing the direct connections between its elements: $A_{ij}$ represents, if it exists, the edge $\{i,j\}$. A specific type of graphs, weighted graphs, are those in which each edge has a numerical weight. In theses cases, $\forall i,j \in V$, the element $A_{ij}$ shows the weight of the edge $e = \{i,j\}$. Note that this weight is not negative, but it does not need to be 0 or 1.

Fuzzy graphs, introduced by Rosenfeld [21], depict other type of networks with a wider scope. These graphs are really useful for modeling situations in which there is some uncertainty. Fuzzy graphs are based on the fuzzy relations among the individuals [19], representing the degree of these relations. Here we provide a formal definition of fuzzy graphs.

**Definition 3. Fuzzy graph** [21]. Let $V$ be a nonempty set, and let be the functions $\eta : V \to [0,1]$, and $\varphi : V \times V \to [0,1]$ such that for all $x,y \in V, \varphi(x,y) \leqslant B(\eta(x), \eta(y))$, where $B$ is a conjunction operator. Then the triplet $G = (V, \eta, \varphi)$ is a fuzzy graph, where $\eta$ is called the fuzzy vertex set, and $\varphi$ is called the fuzzy edge set of $G$.

Assuming that the vertex set $\varphi$ is crisp, previous definition of a fuzzy graph is sometimes simplified, considering that all the information is provided by the pair $G = (V, \varphi)$. In the context of fuzzy graphs, there is another classical assumption about the existence of a crisp set of edges, $E$. Considering this hypothesis, that crisp set of edges limits the value of each fuzzy edge, forcing it to be 0 if the related edge does not exist in $E$. On these assumptions, we propose another definition of a fuzzy graph:

**Definition 4. Crisp graph with fuzzy edges** [22]. Let $G = (V, E)$ be a graph, and let the function $\varphi : E \to [0,1]$ be a fuzzy set defined over the crisp set of edges, $E$. The triplet $\tilde{G} = (V, E, \varphi)$ is named crisp graph $G$ with fuzzy edges $\varphi$.

Let us note that a pair of nodes which are not adjacent in the crisp graph $G$, cannot have any membership degree in $\varphi$. Thus, the available information is just given by the crisp graph $G$ and the weight associated with each edge. Hence, fuzzy graphs can be understood as weighted graphs. Moreover, this structure of a fuzzy graph is equivalent to the structure of the classical FCMs. Thus, as a generalization of fuzzy graphs, we propose the use of extended fuzzy graphs [12]. To set the basis of fuzzy graphs, let us recall the definition of fuzzy measures.

**Definition 5. Fuzzy measure** [23]. Let be the finite set $V$. The set function $\mu : 2^V \to [0,1]$ is a fuzzy measure if the following holds: $\mu(\emptyset) = 0; \mu(V) = 1$; and $\forall A, B \subseteq V$, if $A \subseteq B$, then $\mu(A) \leqslant \mu(B)$.

Hence, an extended fuzzy graph is a graph together with a fuzzy measure. Formally:

**Definition 6. Extended fuzzy graph** [12]. Let $G = (V, E)$ be a graph, and let the function $\mu : 2^V \to [0,1]$ be a fuzzy measure defined over the set of nodes, $V$. The triplet $\tilde{G} = (V, E, \mu)$ is called extended fuzzy graph or crisp graph $G$ with fuzzy measure $\mu$.

Thus far, we have just considered fuzzy measures with an individual interpretation. Now, let us extend the application framework by considering bipolar fuzzy measures [24].

**Definition 7. Bipolar Fuzzy Measure** [24] Let $V$ be a finite set and let $\mu^-, \mu^+ : 2^V \to [0,1]$ be two fuzzy measures. Then, $\mu^b = (\mu^-, \mu^+) : 2^V \times 2^V \to [0,1]^2$ is said to be a bipolar fuzzy measure.

To adapt the interpretation of $\mu^b$ to the purposes of this paper, which are supposed to be based on bipolarity, we assume that $\mu^-$ defines a negative evidence about the elements of $V$, whereas $\mu^+$ defines a positive evidence about these individuals. Moreover, we work under the assumption that the degree of relation among two nodes is a fuzzy relation [25] represented by the fuzzy set $\mu^b$ [26,49]. Owing to the bipolarity of $\mu^b$ that provides a double perception of the interactions, some discrepancy among the nodes is defined by $\mu^-$ and some affinity among the nodes is defined by $\mu^+$. Then, we generalize the notion of the extended fuzzy graph to a wider context based on bipolarity, so we introduce the definition of extended bipolar fuzzy graph.

**Definition 8. Extended bipolar fuzzy graph** [7]. Let be the graph $G = (V, E)$ and let $\mu^b = (\mu^-, \mu^+) : 2^V \times 2^V \to [0,1]^2$ be a bipolar fuzzy measure defined over the set of nodes. The triplet $\tilde{G} = (V, E, \mu^b)$ is an extended bipolar fuzzy graph.

## 2.3. Defining Groups from Fuzzy Measures: The Associated Weighted Graph

Given a fuzzy measure $\mu$ defined over a finite set of objects $X$, it is possible to build a graph whose edges represent the tendency (defined by $\mu$) that have the items of $X$ to stay together. In [12], it was provided a modification of the Louvain algorithm to deal with community detection solution for extended fuzzy graphs. To carry on with that, a key concept was introduced in that paper: the associated weighted graph obtained from a fuzzy measure.

There exist different ways to build a weighted graph from a fuzzy measure. In the proposal done in [12], the definition of this weighted graph was based on the Shapley value related to a fuzzy measure.

**Definition 9. Shapley value** [27]. Let be the fuzzy measure $\mu : V \to [0, 1]$, where $|V| = n$. For every $i \in V$ its Shapley value is calculated as

$$Sh_i(\mu) = \sum_{K \subseteq V \setminus \{i\}} \frac{(n - |K| - 1)! \, |K|!}{n!} (\mu(K \cup \{i\}) - \mu(K))$$

The weighted graph associated with a fuzzy measure [12] is that whose adjacency matrix is defined as

$$\mathcal{F}^\mu_{ij} = \phi \left( Sh_i(\mu) - Sh^j_i(\mu), Sh_j(\mu) - Sh^i_j(\mu) \right) \quad (1)$$

where $\phi : [0, 1]^2 \to [0, 1]$ is bivariate AO; $Sh_i(\mu)$ is the Shapley value of node $i$ in coalition with all the nodes depending on their relation in $\mu$, and $Sh^j_i(\mu)$ is the Shapley value of node $i$ in coalition with all the nodes except the node $j$, depending on their relation in $\mu$.

Then, for each pair of nodes $\{i, j\}$, the weight $\mathcal{F}^\mu_{ij}$ measures how each node is affected by the absence of the other, depending on their relation in $\mu$.

Taking into account that a bipolar fuzzy measure, $\mu^b$, can be viewed as a set of two individual fuzzy measures, $\mu^-$ and $\mu^+$, in [7] it was defined the weighted multi-graph associated with a fuzzy bipolar measure as follows: for each pair of nodes $\{i, j\}$, the weight $\mathcal{F}^-_{ij}$ measures how each node is affected by the absence of the other, depending on their relation in $\mu^-$. Analogously, the weight $\mathcal{F}^+_{ij}$ is about the relation of $i$ and $j$ in $\mu^+$. Hence, matrices $\mathcal{F}^-$ and $\mathcal{F}^+$ are given by Equations (2) and (3).

$$\mathcal{F}^-_{ij} = \phi^- \left( Sh_i(\mu^-) - Sh^j_i(\mu^-), Sh_j(\mu^-) - Sh^i_j(\mu^-) \right) \quad (2)$$

$$\mathcal{F}^+_{ij} = \phi^+ \left( Sh_i(\mu^+) - Sh^j_i(\mu^+), Sh_j(\mu^+) - Sh^i_j(\mu^+) \right) \quad (3)$$

where $\phi^-, \phi^+ : [0, 1]^2 \to [0, 1]$ are two bivariate AOs [28] used to make symmetric the relation between $i$ and $j$ from theirs asymmetrical relations in $\mu^-$ and $\mu^+$; $Sh_i(\mu^-)$ is the Shapley value of node $i$ in coalition with all the nodes depending on their relation in $\mu^-$, and $Sh^j_i(\mu^-)$ is the Shapley value of node $i$ in coalition with all the nodes except the node $j$, depending on their relation in $\mu^-$. It is analogous for $\mu^+$ with $Sh_i(\mu^+)$ and $Sh^j_i(\mu^+)$.

**Definition 10. Bipolar weighted multi-graph associated with a bipolar fuzzy measure.** Let $\mu^b = (\mu^-, \mu^+) : 2^V \times 2^V \to [0, 1]^2$ be a bipolar fuzzy measure. According to the notation previously introduced, let matrices $\mathcal{F}^-$ and $\mathcal{F}^+$ be as follows:

$$\mathcal{F}^-_{ij} = \phi^- \left( Sh_i(\mu^-) - Sh^j_i(\mu^-), Sh_j(\mu^-) - Sh^i_j(\mu^-) \right)$$

$$\mathcal{F}^+_{ij} = \phi^+ \left( Sh_i(\mu^+) - Sh^j_i(\mu^+), Sh_j(\mu^+) - Sh^i_j(\mu^+) \right)$$

$\mathcal{F}^-$ ($\mathcal{F}^+$) is the adjacency matrix of the weighted graph associated with $\mu^-$ ($\mu^+$, respectively). So, the bipolar weighted multi-graph associated with $\mu^b$ is that whose adjacency matrices are $(\mathcal{F}^-, \mathcal{F}^+)$.

## 3. MULTIPLE BIPOLAR FUZZY MEASURES IN GRAPHS

As happen in any statistical real problem, more than one variable is needed to represent in a more natural way the reality. Objects cannot be described just by one variable; there are even many situations in which the similarity (or dissimilarity) between objects depends on more than one criterion. Considering this fact, it is fair to assume that the fuzzy measures that represent the synergies between objects that we are going to cluster, could also not be unidimensional. In addition to this fact, as it was pointed in [7], bipolarity appears in a natural way: there are some capacities that represent the synergies or similarity between objects, but there are also others that represent the dissimilarity or negative synergies (antagonism). Considering all this, in this section we propose a generalization of the extended bipolar fuzzy graph into a multidimensional case and we propose an aggregation process to deal with this complex information.

**Definition 11. Extended multiple bipolar fuzzy graph.** Let $G = (V, E)$ be a graph. For every $\ell \in \{1, \ldots, s\}$, let $\mu^{b_\ell} : 2^V \times 2^V \to [0, 1]^2$ be a bipolar fuzzy measure defined over the set of nodes, $V$. Then, the tuple $\tilde{G} = (V, E, (\mu^{b_1}, \ldots, \mu^{b_s}))$ is called extended multiple bipolar fuzzy graph.

Once the extended multiple bipolar fuzzy graph is defined, we propose a five-step methodology to transform the information provided by the extended multiple bipolar fuzzy graph, for further combination of this transformed information with a community detection algorithm. The process is quite similar to that followed in Definition 10, but in this case, we have to manage $s$ bipolar fuzzy measures.

The first step is to summarize the information given by each bipolar fuzzy measure into two matrices. Particularly, for the bipolar fuzzy measure $\mu^{b_\ell}$, let us denote as $(\mathcal{F}^{-\ell}, \mathcal{F}^{+\ell})$ the adjacency matrices of its associated bipolar weighted multi-graph, obtained according to Equations (2) and (3). Then, the bipolar weighted multi-graph associated with multiple bipolar fuzzy measures is characterized below.

**Definition 12. Bipolar weighted multi-graph associated with multiple bipolar fuzzy measures.** Let $V$ be a finite set. For every $\ell \in \{1, \ldots, s\}$, let $\mu^{b_\ell} : 2^V \times 2^V \to [0, 1]^2$ be a bipolar fuzzy measure, whose associated bipolar weighted multi-graph has $(\mathcal{F}^{-\ell}, \mathcal{F}^{+\ell})$ as adjacency matrices. Then, the bipolar weighted multi-graph associated with the tuple $(\mu^{b_1}, \ldots \mu^{b_s})$ is that whose adjacency matrices are $(\mathcal{F}^{-1}, \ldots, \mathcal{F}^{-s}, \mathcal{F}^{+1}, \ldots, \mathcal{F}^{+s})$.

This new concept (bipolar weighted multi-graph associated with multiple bipolar fuzzy measures) summarizes (particularly for our clustering purpose) the information provided by the $s$ bipolar fuzzy measures. To deal with this multidimensional problem, we propose to aggregate all the information defined by the bipolar weighted multi-graph associated with multiple bipolar fuzzy measures. In particular, we propose a natural extension of this idea, which was firstly introduced in [7], to a multidimensional bipolar case, in which there are many fuzzy positive measures and many negative ones. On the one hand, we aggregate all the positive information; on the other hand, we aggregate the negative one. We consider that groups/communities are formed by those nodes that present high positive tendency and low negative one. Below we summarize the five-step process used to summarize multiple bipolar fuzzy measures into a final matrix, whose final aim is devoted to community detection problems.

- **Step 1: Built bipolar weighted multi-graph.** Given a multiple extended bipolar fuzzy graph, the adjacency matrices $\left(\mathcal{F}^{-1}, \ldots, \mathcal{F}^{-s}, \mathcal{F}^{+1}, \ldots, \mathcal{F}^{+s}\right)$ are obtained.

- **Step 2: Aggregating negative information**. In this case, we first aggregate matrices $\mathcal{F}^{-1}, \ldots, \mathcal{F}^{-s}$ into matrix $\mathcal{F}^-$. To do that, we use an aggregation function $\Phi^- : \mathcal{X}^s \to \mathcal{X}$ defined over the set of quadratic $n$-matrices $\mathcal{X}$. In this paper, we propose the use of matrix aggregation based on classical aggregation with a point-wise transformation. Formally, we have

$$\Phi^-\left(\mathcal{F}^{-1}, \ldots, \mathcal{F}^{-s}\right) = \mathcal{F}^-$$

where

$$(\mathcal{F}^-)_{ij} = A_s\left((\mathcal{F}^{-1})_{ij}, \ldots, (\mathcal{F}^{-s})_{ij}\right)$$

being $A_s$ a classical AO (see Subsection 2.1).

- **Step 3: Aggregating positive information**. In a similar way, we aggregate the matrices $\mathcal{F}^{+1}, \ldots, \mathcal{F}^{+s}$ into matrix $\mathcal{F}^+$, by means of the aggregation function $\Phi^+ : \mathcal{X}^s \to \mathcal{X}$. As we have done before, we propose the use of matrix aggregation functions based on classical AOs. Formally, we have

$$\Phi^+\left(\mathcal{F}^{+1}, \ldots, \mathcal{F}^{+s}\right) = \mathcal{F}^+$$

where

$$(\mathcal{F}^+)_{ij} = B_s\left((\mathcal{F}^{+1})_{ij}, \ldots, (\mathcal{F}^{+s})_{ij}\right)$$

being $B_s$ a classical AO.

- **Step 4: Transforming negative information.** Let us note that matrices $\mathcal{F}^-$ and $\mathcal{F}^+$ have opposite meanings (one is about discrepancy and the other is about affinity). Then, to obtain a cohesive aggregation of them, in this step we propose a transformation of the matrix $\mathcal{F}^-$ into its "antonym," $\mathcal{F}^-_{op}$ (this antonym may have several interpretations). Let $N : \mathcal{X} \to \mathcal{X}$ be a negation transformation. Then, we define $\mathcal{F}^-_{op} = N(\mathcal{F}^-)$. $N$ should be chosen so that matrices $\mathcal{F}^+$ and $\mathcal{F}^-_{op}$ have the same nature.

- **Step 5: Final aggregation**. Finally, we combine the information provide by $\mathcal{F}^-_{op}$ and $\mathcal{F}^+$ by means of the AO $\psi : \mathcal{X}^2 \to \mathcal{X}$ as follows:

$$\mathcal{F}^*_b = \psi\left(\mathcal{F}^-_{op}, \mathcal{F}^+\right)$$

where

$$(\mathcal{F}^*_b)_{ij} = C_2\left((\mathcal{F}^-_{op})_{ij}, (\mathcal{F}^+)_{ij}\right)$$

being $C_2$ a classical bi-variate AO.

**Remark 1.** Note that previous process involves three different classes of AOs: one for positive information, other for the negative one, and finally, another for combining the positive and negative information. In this paper we present the aggregation problem/process in a general way since any AO could be used. We think that the aggregation should be strongly dependent on the real problem that you are dealing with.

In the unidimensional case, we only used this last class of AO, so just three notions of group were contemplated:

- If $C_2$ is disjunctive, then the elements of the groups searched will have positive relations or they will not have negative relations.

- If $C_2$ is conjunctive, then the elements of the groups searched will have positive relations and they will not have negative relations.

- If $C_2$ is an averaging operator, we will contemplate an equilibrium among both previous points.

The problem becomes more complex when there are several information sources, with positive and negative meanings, which have to be aggregated. If we force to the AOs $A_s$, $B_s$, and $C_2$ to belong to the first three groups defining in Subsection 2.1 we will have 27 different types of groups. The two extreme cases are the following:

- $A_s$, $B_s$, and $C_2$ are disjunctive. In this the case, all the groups whose individuals share (positively or negatively) some common characteristic, will be enhanced. Obviously, as we increase the number of bipolar measures, $s$, groups will be bigger.

- $A_s$, $B_s$, and $C_2$ are conjunctives. This case is opposite to the previous one. Here, the only enhanced groups are those that have all the positive characteristics in common, and which have no tendency to be separated into any negative measure. Obviously, as we increase the number of bipolar measures, $s$, the groups will be more restrictive.

In the computational results section, we have used average AOs for $A_s$ and $B_s$ and conjunctive AO for $C_2$.

## 4. THE MULTIPLE BIPOLAR DUO LOUVAIN ALGORITHM

This section is based on one of the most popular algorithms in community detection framework: the Louvain algorithm [15]. Hence,

first we explain the main points of this method, to provide a clear vision of its performance. Then, we propose a modification of it that allows us to deal with problems in which there is some additional information. Particularly, we propose an application of this modification to deal with community detection problems based on multiple bipolar fuzzy measures.

## 4.1. Louvain Algorithm

In last decades, the interest of many researchers has been focused on community detection problem, so many effective method have been proposed in this field. In 2008, Blondel *et al.* introduced one of the most popular algorithms in networks literature: the Louvain algorithm [15]. Based on local moving heuristic and modularity optimization [29], this algorithm, widely used due to its speed and effectiveness, provides a nonhierarchical partition, obtained in a multi-phase process. Two concepts are essential on this process: the modularity of a partition and the variation of modularity obtained when moving node $i$ to $C_j$ ($j$'s community), denoted as $\Delta Q_i(j)$.

**Definition 13. Modularity** $Q$ [30] Let $G = (V, E)$ be a graph, where $m = |E|$. Let $P$ be a partition of the nodes. The modularity of $P$ is calculated as

$$Q = \frac{1}{2m} \sum_{i,j \in V} \left[ A(i,j) - \frac{k_i k_j}{2m} \right] \delta(C_i, C_j)$$

being $A$ the adjacency matrix of the graph, $k_i$ the degree of the node $i$, and $C_i$ the cluster to which node $i$ belongs. The value of $\delta(C_i, C_j)$ is 1 if nodes $i$ and $j$ are in the same cluster, and 0 otherwise.

**Definition 14. Variation of modularity** $\Delta Q_i(j)$ [15]. Let $G = (V, E)$ be a graph, where $i, j \in V$. Having a partition of its nodes, let $C_j$ be the group to which $j$ belongs. Let $\sum_{in}$ be the sum of the weights of the edges inside $C_j$, and let $\sum_{tot}$ be the sum of the weights of the edges incident to nodes in $C_j$. $k_i$ is the degree of node $i$; $k_{i,in}$ is the number of edges from $i$ to nodes in $C_j$, and $m = |E|$. The variation of modularity obtained when moving node $i$ to $C_j$ is calculated as

$$\Delta Q_i(j) = \left[ \frac{\sum_{in} + 2k_{i,in}}{2m} - \left( \frac{\sum_{tot} + k_i}{2m} \right)^2 \right]$$
$$- \left[ \frac{\sum_{in}}{2m} - \left( \frac{\sum_{tot}}{2m} \right)^2 - \left( \frac{k_i}{2m} \right)^2 \right]$$

Now we show a sketch about Louvain performance, which is divided into two phases.

- **Phase 1.** At the beginning, each node of the graph defines in itself an isolated community. For each node $i$ and all its neighbors $j$, we calculate $\Delta Q_i(j)$, so that $j^*$ is the node for which the value $\Delta Q_i$ is maximum. If $\Delta Q_i(j^*) > 0$, then node $i$ is moved to the community to which $j^*$ belongs. Else, $i$ remains in its group. This process is sequentially and repeatedly applied for the entire set of nodes with random order (in fact, one node could be analyzed several times), until a local maximum of the modularity is reached. Then, the first phase ends.

- **Phase 2.** Considering the communities defined in the first phase as nodes, which are called supervertices [31], the second phase starts with the construction of a new network whose nodes are these supervertices. Two supervertices are connected if there is at least one edge among them, i.e., there is at least one node in one supervertice connected with one node in the other supervertice. To calculate the weight of each supervertice, we have to sum the weight of the edges between the communities found in the original graph. Furthermore, there could be self-loops, given by the edges between nodes which are in the same supervertice. Once the new network is defined, the next part of this phase consists on the application of the first phase, considering the new graph.

Both phases are iteratively repeated while there are changes in modularity, until a maximum of modularity is achieved. In Algorithm 1 we provide a sketch of the performance of this algorithm, by means of its pseudocode.

Note that, because of its order dependence, the solution obtained with the Louvain algorithm is not unique [32]. Usually, several outputs of the same networks are analyzed, considering random reshuffling of the order in which the nodes are evaluated [33].

---

**Algorithm 1:** *Louvain* input= A output= P
---
1: **Phase 1**.
2: $o = permutation(V)$
3: Let each node of the graph be an isolated community
4: **While** There is some change in modularity optimization
5:    According to the order given by $o$, let $i$ be the corresponding element. Then, find all $j$ neighbour of $i$ in $A$
6:    Calculate $\Delta Q_i(j)$ in matrix $A$
7:    Let $j^*$ be the node for which $\Delta Q_i$ is maximum
8:    **If** $\Delta Q_i(j^*) > 0$
9:      Move node $i$ to the community to which $j^*$ belongs
10:    **Else**
11:      $i$ remains in its community
12: **End while**
13: **Phase 1 Ends**
14: **Phase 2**
15:    $A^*$ is the aggregated matrix obtained from $A$, whose nodes are the communities found in Phase 1
16:    While there is some change, apply the Louvain Algorithm, considering matrix $A^*$
17: **Phase 2 Ends**

---

## 4.2. Louvain Algorithm Over Extended Multiple Bipolar Fuzzy Graphs

In this section we explain a modification of the Louvain algorithm [15] to deal with situations in which there is some additional information about the relations among the nodes, given by multiple bipolar fuzzy measures which are independent of the structure of the graph.

First, we propose a new concept of group which is based on the notion of extended multiple bipolar fuzzy graph (see Definition 11). Hence, let $\tilde{G} = \left(V, E, \left(\mu^{b_1}, \ldots, \mu^{b_s}\right)\right)$ be an extended multiple bipolar fuzzy graph in which, $\forall \ell \in \{1, \ldots, s\}$, $\mu^{b_\ell} = \left(\mu^{-\ell}, \mu^{+\ell}\right)$ is a bipolar fuzzy measure, where fuzzy measures $\mu^{-\ell}$ and $\mu^{+\ell}$ define some discrepancy/affinity relations among the elements of $V$. As we have detailed in previous section, three operators are needed to define the matrix $\mathcal{F}_b^*$, which summarizes all the information encoded in the multiple bipolar fuzzy measures $\mu^{b_1}, \ldots, \mu^{b_s}$, apart from the functions $\phi^{-\ell}, \phi^{+\ell}$ needed to define each individual $\mathcal{F}^{-\ell}, \mathcal{F}^{+\ell}$. These three operators are $\Phi^-$ and $\Phi^+$, to aggregate all the negative and positive information into individuals matrices, and, $\psi$, to combine all the information into the final single matrix, $\mathcal{F}_b^*$. Hence, the notion of a group will depend on the choice about these three operators. Particularly, we focus on the consideration of the ordered weighted averaging (OWA) AOs [34–37]. The most notable OWA operators are the maximum, the minimum, and the average. In this background of clustering problems with multiple bipolar fuzzy measures, in which the notion of a group depends on the selection made for each operator, $\Phi^-, \Phi^+$, and $\psi$, the possible permutations of choices done for these operators provide 27 different concepts of group. For example, if we consider $\Phi^- = \Phi^+ = max$, and $\psi = min$ we will be looking for groups whose nodes have strong affinity relations, and between which there are weak, even nonexistent discrepancy relations. Other case, if there is some interest about a balance of the whole information, in order to make up for the positive and the negative evidences, we could consider $\Phi^- = \Phi^+ = \psi = average$. Hence, several permutations of OWA operators provide us diferent notions of groups. This selection could be adapted to the problem addressed.

On the other hand, to carry on with community detection problems with additional information given by multiple bipolar fuzzy measures, let us propose an alternative vision of the Louvain algorithm. We want to differentiate two input parameters in this algorithm. One, the matrix used to find "possible" clusters. It provides information about the edges between the nodes. The other is the matrix used to calculate the modularity optimization for a possible local moving. Obviously, the Louvain algorithm considers the same matrix for both purposes (the adjacency matrix of the graph).

Now we define an algorithm based on this alternative vision of the Louvain algorithm. Let $A$ be the adjacency matrix of the graph $G = (V, E)$, and let $\mathcal{M}$ be some matrix. The definition of algorithm *Duo Louvain* is summarized by the following pseudocode, being $\Delta Q_i(j)$ the variation of modularity obtained when moving the node $i$ to $j$'community.

Let us remark that the complexity of *Duo Louvain* algorithm is exactly the same as the complexity of the Louvain algorithm.

Then we explain how to apply the proposed algorithm when solving community detection problems based on multiple bipolar fuzzy measures. Let $G = (V, E)$ be a graph, and, for every $\ell = 1, \ldots, s$, let $\mu^{b_\ell} = (\mu^{-\ell}, \mu^{+\ell}) : 2^V \times 2^V \to [0, 1]^2$ be a bipolar fuzzy measure defined over the set of nodes, $V$, so that the tuple $\tilde{G} = (V, E, (\mu^{b_1}, \ldots, \mu^{b_s}))$ is an extended multiple bipolar fuzzy graph.

To find communities in $\tilde{G}$, first we have to calculate the bipolar weighted multi-graph associated with $(\mu^{b_1}, \ldots, \mu^{b_s})$ (see Definition 12). Once its adjacency matrix $\mathcal{F}_b^*$ is obtained, we apply the Algorithm 2 with input parameters $A$ (the adjacency matrix of $G$), and $\mathcal{M} = \theta(A, \mathcal{F}_b^*)$, being $\theta : \mathcal{X}^2 \to \mathcal{X}$ a function used to combine several matrices.

We will explain with a pseudocode this particular application of Algorithm 2 for solving community detection problems based on multiple bipolar fuzzy measures. For each bipolar fuzzy measure

**Algorithm 2:** *Duo Louvain* input= $(\mathbf{A}, \mathcal{M})$ output= $\mathbf{P}$

1: **Phase 1**.
2: $o = permutation(V)$
3: Let each node of the graph be an isolated community
4: **While** There is some change in modularity optimization
5: According to the order given by $o$, let $i$ be the corresponding element. Then, find all $j$ neighbour of $i$ in $A$
6: Calculate $\Delta Q_i(j)$ in matrix $\mathcal{M}$
7: Let $j^*$ be the node for which $\Delta Q_i$ is maximum
8: **If** $\Delta Q_i(j^*) > 0$
9: Move node $i$ to the community to which $j^*$ belongs
10: **Else**
11: $i$ remains in its community
12: **End while**
13: **Phase 1 Ends**
14: **Phase 2**
15: $A^*$ is the aggregated matrix obtained from $A$, whose nodes are the communities found in Phase 1
16: $\mathcal{M}^*$ is the aggregated matrix obtained from $\mathcal{M}$, whose nodes are the communities found in Phase 1
17: While there is some change, apply the Duo Louvain Algorithm, considering matrices $A^*$ to find edges and $\mathcal{M}^*$ to modularity optimization
18: **Phase 2 Ends**

$\mu^{b_\ell}$, we make use of the aggregation functions $\phi^{-\ell} : [0, 1]^2 \to [0, 1]$, $\phi^{+\ell} : [0, 1]^2 \to [0, 1]$, both used to make symmetric the relation between nodes $i$ and $j$ from theirs asymmetrical relations in $\mu^{-\ell}$ and $\mu^{+\ell}$, respectively. After that, being $\mathcal{X}$ the set of quadratic $n$-matrices, the aggregation functions $\Phi^- : \mathcal{X}^s \to \mathcal{X}$ and $\Phi^+ : \mathcal{X}^s \to \mathcal{X}$ are used to aggregate the negative and the positive information into matrices $\mathcal{F}^-$ and $\mathcal{F}^+$, respectively. Then, we combine the negative and positive information about the relations among the nodes given by $\mathcal{F}^-$ and $\mathcal{F}^-$ with $\psi : \mathcal{X}^2 \to \mathcal{X}$. Finally, we combine the different information sources we have (the graph, $A$, and the additional information, $\mathcal{M}$) with $\theta : \mathcal{X}^2 \to \mathcal{X}$. Let us recall that the notion of a group will depend on $\Phi^-, \Phi^+$, and $\psi$. Let $N : \mathcal{X} \to \mathcal{X}$ be a transformation of $\mathcal{F}^-$ into its antonym, $\mathcal{F}_{op}^-$. The *Multiple Bipolar Duo Louvain* Algorithm is summarized below in Algorithm 3.

**Algorithm 3:** *Multiple Bipolar Duo Louvain* input = $(\mathbf{A}, (\mu^{\mathbf{b_1}}, \ldots, \mu^{\mathbf{b_s}}))$ output = $\mathbf{P}$

1: **for** $\ell = 1 : s$ **do**
2: $\mathcal{F}_{ij}^{-\ell} = \phi^{-\ell}\left(Sh_i(\mu^{-\ell}) - Sh_i^j(\mu^{-\ell}), Sh_j(\mu^{-\ell}) - Sh_j^i(\mu^{-\ell})\right)$
3: $\mathcal{F}_{ij}^{+\ell} = \phi^{+l}\left(Sh_i(\mu^{+\ell}) - Sh_i^j(\mu^{+\ell}), Sh_j(\mu^{+\ell}) - Sh_j^i(\mu^{+\ell})\right)$
4: **end for**
5: $\mathcal{F}^- = \Phi^-\left(\mathcal{F}^{-1}, \ldots, \mathcal{F}^{-s}\right)$
6: $\mathcal{F}^+ = \Phi^+\left(\mathcal{F}^{+1}, \ldots, \mathcal{F}^{+s}\right)$
7: $\mathcal{F}_{op}^- = N(\mathcal{F}^-)$
8: $\mathcal{F}_b^* = \psi\left(\mathcal{F}_{op}^-, \mathcal{F}^+\right)$
9: $\mathcal{M} = \theta\left(A, \mathcal{F}_b^*\right)$
10: Apply *Duo Louvain* $(A, \mathcal{M})$

Regarding the calculation of Shapley value, it has an exponential complexity, which can be partially solved with sampling techniques

[38,39]. However, we want to remark that when additive fuzzy measures are considered, the exact calculation of this value is immediate [12]. Hence, in this context, the complexity of Algorithm 3 is exactly the same as the Louvain algorithm.

Let us recall the example proposed in the Introduction. Now we show that the Algorithm 3, *Multiple Bipolar Duo Louvain* finds the communities which seemed to be logical when considering the affinity/discrepancy relations among the individuals.

**Example 1.** Let $G = (V, E)$ be the graph showed in the Introduction (Figure 1). Then, let $\tilde{G} = (V, E, (\mu^{b_1}, \mu^{b_2}))$ be the extended multiple bipolar fuzzy graph, whose related crisp graph has as adjacency matrix that showed in Figure 2a. Regarding the personal/working relationships proposed in the Introduction, let $\mathcal{F}^{-1}$ and $\mathcal{F}^{-2}$ (Figure 2b and 2d) be the matrices representing the discrepancy relations between the individuals, according to the personal relationships and the working relations, respectively. Let $\mathcal{F}^{+1}$ and $\mathcal{F}^{+2}$ (Figure 2c and 2e) be the matrices representing the affinity among the individuals depending on their personal relationships or their working relations, respectively. Matrices in Figure 2 represents the evolution from the crisp graph provided in the Introduction (Figure 1) to an extended multiple bipolar fuzzy graph. Let us note that the information provided by the multiple bipolar fuzzy measures, $\mu^b = (\mu^{b_1}, \mu^{b_2})$ goes far away from what could be defined by means of a fuzzy graph. For example, considering this additional information, we know that elements $2 - 4$ have a "bad" relation. However, in the crisp graph $G$, there is no edge in those nodes, so this information cannot be represented with a fuzzy graph.

We calculate $\mathcal{F}^- = \Phi^- (\mathcal{F}^{-1}, \mathcal{F}^{-2}) = \max\{\mathcal{F}^{-1}, \mathcal{F}^{-2}\}$ and $\mathcal{F}^+ = \Phi^+ (\mathcal{F}^{+1}, \mathcal{F}^{+2}) = \max\{\mathcal{F}^{+1}, \mathcal{F}^{+2}\}$. Considering $\mathcal{F}_b^* = \psi (\mathcal{F}_{op}^-, \mathcal{F}^+) = \min\{\mathcal{F}_{op}^-, \mathcal{F}^+\}$, where $\mathcal{F}_{op}^- = N(\mathcal{F}^-) = 1 - \mathcal{F}^-$ and $\theta (A, \mathcal{F}_b^*) = \frac{1}{2}A + \frac{1}{2}\mathcal{F}_b^*$, the partition obtained with the application of the Algorithm 3, *Multiple Bipolar Duo Louvain*, is $\mathcal{P}^a = \{\{1, 2\}, \{3, 4\}, \{5, 6\}, \{7, 8\}\}$. However, if we just consider the topology of the graph, the partition obtained with the Louvain algorithm is $\mathcal{P} = \{\{1, 2, 3, 4\}, \{5, 6, 7, 8\}\}$.

## 5. COMPUTATIONAL RESULTS

In this section we provide some computational results in order to test the effectiveness of the Algorithm 2, with the particular application given in Algorithm 3 used to solve community detection problems based on multiple bipolar fuzzy measures. To carry on with it, we work with benchmark models [16]. Here we propose four different "standard" or "gold" structures. The objective is to quantify, by means of the *NMI*) [17], how much information of this fixed communities structure recovers our algorithm. Let us give a short explanation about the calculation of *NMI*.

**Definition 15.** *Normalized Mutual Information NMI* [17] Let $X = \{x_i\}_{i \in V}$ and $Y = \{y_i\}_{i \in V}$ be two disjoint partitions of the graph $G = (V, E)$. We denote as $P(x)$ the probability that a random node is assigned to community $x$, and we denote as $P(x, y)$ the conditioned probability that a node is simultaneously assigned to $x$ by $X$ and to $y$ by $Y$. So, the Shanon entropy for $X$ is calculated as $H(X) = -\sum_x P(x) log(P(x))$ and the Shanon entropy for $X$ and $Y$ is calculated as $H(X, Y) = -\sum_x \sum_y P(x, y) log(P(x, y))$. The mutual information between $X$ and $Y$, *MI*, is calculated as

$$MI(X, Y) = \sum_x \sum_y P(x, y) \log \frac{P(x, y)}{P(x) P(y)} \quad (4)$$

Then, the *NMI* is just a normalization of the *MI*.

$$NMI(X, Y) = \frac{2MI(X, Y)}{H(X) + H(Y)} \quad (5)$$

According to Equation (5), the *NMI* shows how similar two clusters are: if *NMI* = 1, then both clusters are totally equal, whereas if *NMI* = 0, they are totally dissimilar.

### 5.1. Benchmark Graphs

When a new method is proposed, one of the most important issues is its comparison with the existing work, in order to test if the new proposal is at least as good as the old ones. One crucial point when testing clustering methods is about which evidence or characteristic should be analyzed: not only the complexity or speed of the method should be considered but also the goodness of the results should be measured. The issue is that each method has its own characteristics and peculiarities, even the output might not be comparable. So, in this work we focus the test on benchmarking. In [40], Bader *et al.* said "*Benchmarking refers to a repeatable performance evaluation as a means to compare somebody's work to the state of the art in the respective field.*"

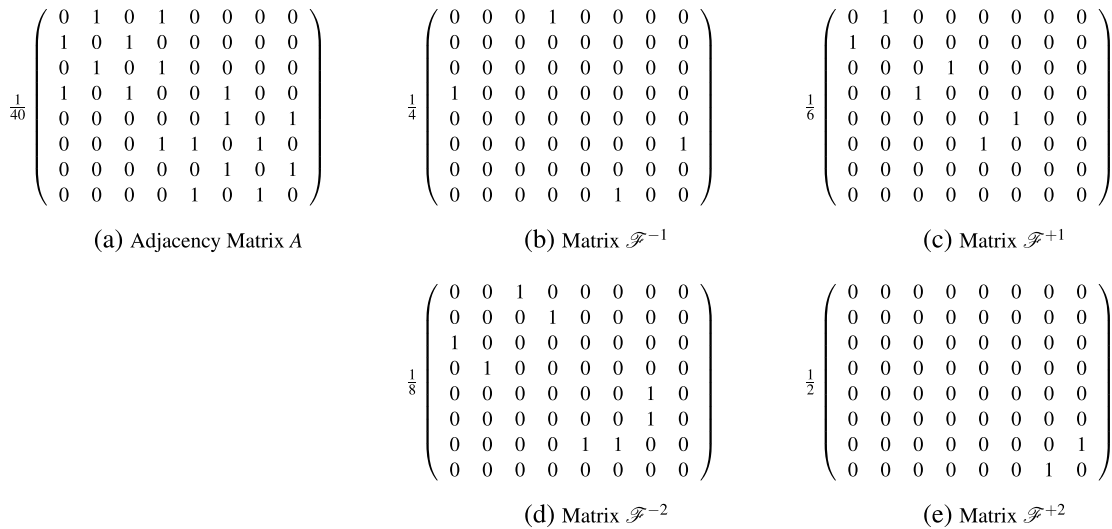

**Figure 2** | Adjacency matrix $A$, and relations matrices $\mathcal{F}^{-1}, \mathcal{F}^{+1}, \mathcal{F}^{-2}, \mathcal{F}^{+2}$.

Then, in this paper we artificially build benchmark graphs. These are synthetic graphs generated with an embedded cluster structure. This structure is assumed to be a "standard" or gold partition, and the main point is to analyze the ability of an algorithm to find it. We quantify this ability with the calculation of the *NMI*: we measure how much the obtained result resembles the "gold" partition [41]. So, the closer to 1 the *NMI* is, the better the result obtained with analyzed algorithm is.

To generate these benchmark models, we assume that both the expected degree, $<k>$, and the community size, $|C_t|$, are power laws, whose exponents are $\alpha$ and $\beta$, respectively. Then, the construction is based on Equation (6): given the cluster $C_t$, the probability that there is an edge between nodes $i$ and $j$ is

$$P(i,j) = \begin{cases} \alpha & \text{if} \quad i,j \in C_k \\ \beta & \text{if} \quad \text{otherwise} \end{cases} \quad (6)$$

## 5.2. Results

In this subsection we generate four different benchmark graphs, based on the idea proposed in [4], as it is explained in the previous section. We work in a broad context, considering also asymmetric structures. Each model, representing an extended multiple bipolar fuzzy graph $\tilde{G} = \left(V, E, \left(\mu^{b_1}, \dots, \mu^{b_s}\right)\right)$, will have two different and independent components. One of them is about the structure of the graph; it is given by the direct connection among the nodes, it is, the edges in $E$. The other part, which is obtained from an aggregation of some bipolar fuzzy measures $\mu^{b_1}, \dots, \mu^{b_s}$ into matrices $\mathcal{F}^-$ and $\mathcal{F}^+$, is about the relations of discrepancy/affinity among different pairs of nodes, respectively.

Both the adjacency matrix $A$ and the relationship matrices $\mathcal{F}^-$ and $\mathcal{F}^+$ have been generated randomly with 256 nodes, considering different values of in/out degree ($z_{in}$ and $z_{out}$, respectively), depending on parameters $\alpha$ y $\beta$ (see Table 1).

**Table 1** | Benchmark parameters.

| Label | $\alpha$ | $\beta$ |
|---|---|---|
| 1 | 0.45 | 0.016 |
| 2 | 0.4 | 0.033 |
| 3 | 0.35 | 0.05 |
| 4 | 0.325 | 0.058 |
| 5 | 0.3 | 0.066 |
| 6 | 0.275 | 0.075 |
| 7 | 0.25 | 0.083 |
| 8 | 0.225 | 0.091 |
| 9 | 0.2 | 0.1 |

In all the cases showed in this section, matrices $\mathcal{F}_{op}^-$ and $\mathcal{F}^+$ are aggregated with the minimum operator, it is, $\mathcal{F}_b^* = \min\{\mathcal{F}_{op}^-, \mathcal{F}^+\}$, being $\mathcal{F}_{op}^- = N(\mathcal{F}^-) = 1 - \mathcal{F}^-$. Note that this definition of matrix $\mathcal{F}_b^*$ holds a notion of a group which should be composed of those elements between which there are high affinity relations and no discrepancy relations. Obviously, any other aggregation function may be selected, so the notion of group will variate.

For each combination of the parameters $\alpha$ and $\beta$, we analyze different linear aggregation of corresponding matrices, $\mathcal{M} = \theta\left(A, \mathcal{F}_b^*\right) = \gamma A + (1 - \gamma)\mathcal{F}_b^*$, considering different values for the aggregation parameter $\gamma$, ($\gamma = 1, 0.75, 0.5, 0.25, 0$). In Tables 2–5, we show an average of the *NMI* resulting from 100 iterations of each combination of the values $\alpha$ and $\beta$ for both networks, generated with 256 nodes, when $\gamma = 0$ (in this case, the results are optimal for all the benchmarks).

### 5.2.1. Benchmark model. Case 1

We analyze the most basic case Figure 3. We consider an extended multiple bipolar fuzzy graph with 256 nodes, whose adjacency matrix has two well-separated groups in its main diagonal. Each of them is supposed to have 128 nodes, where $<k> = 128\alpha + 128\beta$ is the expected *in* degree for each one. Both relations graph are defined with four communities, $C_1, \dots, C_4$, whose expected size is $|C_i| = 64$. The expected *in* degree for each node in the affinity relations graph $\mathcal{F}_1^+$ is $<k> = 64\alpha + 192\beta = 32$, whereas in the discrepancy relations graph $\mathcal{F}_1^-$ is $<k> = 192\alpha + 64\beta$.

In Table 2 we show the results obtained for different values of $\alpha$ and $\beta$. The information about the label assigned to each value of these parameters can be found in Table 1. Let us remark that the goodness of the algorithm does not depend on the graph at all. In fact, it gets worse as the relations matrix becomes scarcer (column by column). Nevertheless, in this case in which the standard partition is symmetric and high enough, the results provided by Algorithm 3 are very good, as for all the cases, the NMI is almost 1.

### 5.2.2. Benchmark model. Case 2

It is known that the Louvain algorithm is very size sensitive. Nevertheless, in this subsection we show that our algorithm also works well when considering smaller communities. Here we define a standard structure in both relations graphs $\mathcal{F}_2^-$ and $\mathcal{F}_2^+$ with 8 communities, each them supposed to have 32 nodes. The expected *in* degree of each node in each community of matrix $\mathcal{F}_2^+$ is $<k> = 32\alpha + 224\beta$, whereas in matrix $\mathcal{F}_2^-$, the expected *in* degree is $<k> = 224\alpha + 32\beta$. The size of the groups in the graph used to find "possible" clusters is also reduced, as it can be seen in Figure 4. In this case, the adjacency matrix $A_2$ is defined with four embedded communities, $C_1, \dots, C_4$, all supposed to have 64 nodes. The expected *in* degree for each node in the connections graph is $<k> = 64\alpha + 192\beta = 32$. Obtained results can be seen in Table 3. It can be seen that, although the size of the communities is reduced, the Algorithm 3 still provides "good" partitions, whose NMI is near 1.

### 5.2.3. Benchmark model. Case 3

The issue when working with previous benchmark models is that both may not fit well with reality, specially because all communities have the same size, and all nodes have the same expected degree. Then, an algorithm may work well in those benchmark graphs, but it is not as good at performing real-life examples. Hence, a good benchmark model should have a skewed degree distribution similar to real networks [42]. In general, algorithms should be tested on benchmarks of variable average degree and community size because these parameters may affect seriously the results. In an attempt to provide a better representation of reality, here we proposed a more complex structure: we consider asymmetric relations matrices $\mathcal{F}_3^-$ and $\mathcal{F}_3^+$, in which communities have different sizes (see Figure 5). Then, the expected size of these 5 clusters is $|C_1| = 43$, $|C_2| = 42$, $|C_3| = 43$, $|C_4| = 96$, $|C_5| = 32$. Regarding the adjacency matrix $A_3$, it is built in the same way as it is built matrix $A_1$ in Subsection 5.2.1. In Table 4 we show that our proposal works very

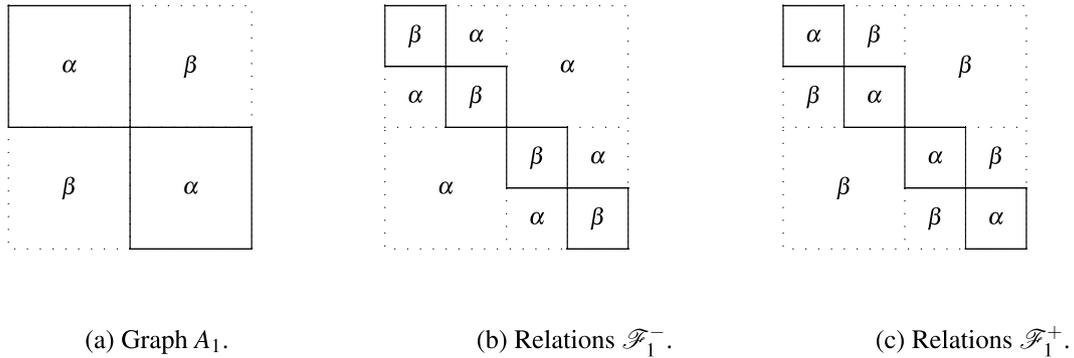

(a) Graph $A_1$.　　(b) Relations $\mathcal{F}_1^-$.　　(c) Relations $\mathcal{F}_1^+$.

**Figure 3** | Benchmark model. Case 1.

**Table 2** | Normalized mutual information (*NMI*) Benchmark model case 1.

| NMI | Relations Value 1 | Relations Value 2 | Relations Value 3 | Relations Value 4 | Relations Value 5 | Relations Value 6 | Relations Value 7 | Relations Value 8 | Relations Value 9 |
|---|---|---|---|---|---|---|---|---|---|
| Graph Value 1 | 1 | 1 | 1 | 1 | 1 | 0.9999 | 0.9985 | 0.9667 | **0.8159** |
| Graph Value 2 | 1 | 1 | 1 | 1 | 1 | 1 | 0.9980 | 0.9712 | **0.8115** |
| Graph Value 3 | 1 | 1 | 1 | 1 | 1 | 1 | 0.9983 | 0.9597 | **0.8056** |
| Graph Value 4 | 1 | 1 | 1 | 1 | 1 | 1 | 0.9970 | 0.9964 | **0.8029** |
| Graph Value 5 | 1 | 1 | 1 | 1 | 1 | 1 | 0.9983 | 0.9640 | **0.8091** |
| Graph Value 6 | 1 | 1 | 1 | 1 | 1 | 1 | 0.9987 | 0.9555 | **0.8087** |
| Graph Value 7 | 1 | 1 | 1 | 1 | 1 | 1 | 0.9978 | 0.9686 | **0.8004** |
| Graph Value 8 | 1 | 1 | 1 | 1 | 1 | 1 | 0.9964 | 0.9591 | **0.8067** |
| Graph Value 9 | 1 | 1 | 1 | 1 | 1 | 1 | 0.9978 | 0.9683 | **0.8091** |

**Table 3** | Normalized mutual information (*NMI*) Benchmark model case 2.

| NMI | Relations Value 1 | Relations Value 2 | Relations Value 3 | Relations Value 4 | Relations Value 5 | Relations Value 6 | Relations Value 7 | Relations Value 8 | Relations Value 9 |
|---|---|---|---|---|---|---|---|---|---|
| Graph Value 1 | 1 | 1 | 1 | 1 | 0.9978 | 0.9891 | 0.9703 | 0.9814 | 0.9939 |
| Graph Value 2 | 1 | 1 | 1 | 1 | 0.9995 | 0.9900 | **0.9801** | 0.9883 | 0.9999 |
| Graph Value 3 | 1 | 1 | 1 | 1 | 0.9981 | 0.9943 | 0.9935 | 0.9919 | 0.9937 |
| Graph Value 4 | 1 | 1 | 1 | 1 | 0.9991 | 0.9948 | 0.9921 | 0.9907 | 0.9997 |
| Graph Value 5 | 1 | 1 | 1 | 1 | 0.9992 | 0.9929 | **0.9877** | 0.9701 | 0.9950 |
| Graph Value 6 | 0.9999 | 1 | 1 | 1 | 0.9998 | 0.9935 | **0.9849** | 0.9882 | 0.9903 |
| Graph Value 7 | 0.9999 | 1 | 1 | 1 | 0.9995 | 0.9981 | 0.9928 | 0.9935 | 0.9917 |
| Graph Value 8 | 0.9998 | 0.9998 | 1 | 1 | 1 | 0.9971 | 0.9950 | **0.9878** | 0.9821 |
| Graph Value 9 | 0.9991 | 0.9997 | 0.9991 | 0.9997 | 0.9994 | 0.9991 | 0.9979 | 0.9899 | 0.9895 |

**Table 4** | Normalized mutual information (*NMI*) Benchmark model case 3.

| NMI | Relations Value 1 | Relations Value 2 | Relations Value 3 | Relations Value 4 | Relations Value 5 | Relations Value 6 | Relations Value 7 | Relations Value 8 | Relations Value 9 |
|---|---|---|---|---|---|---|---|---|---|
| Graph Value 1 | 1 | 1 | 1 | 1 | 0.9995 | 0.9929 | 0.9680 | **0.9145** | 0.8431 |
| Graph Value 2 | 1 | 1 | 1 | 1 | 0.9989 | 0.9916 | 0.9608 | **0.9116** | 0.8405 |
| Graph Value 3 | 1 | 1 | 1 | 1 | 0.9984 | 0.9935 | 0.9627 | **0.9067** | 0.8418 |
| Graph Value 4 | 1 | 1 | 0.9997 | 0.9997 | 0.9989 | 0.9938 | 0.9600 | **0.9045** | 0.8398 |
| Graph Value 5 | 1 | 1 | 1 | 0.9999 | 0.9976 | 0.9859 | 0.9721 | **0.9068** | 0.8431 |
| Graph Value 6 | 1 | 1 | 1 | 0.9987 | 0.9980 | 0.9926 | 0.9615 | **0.8947** | 0.8452 |
| Graph Value 7 | 1 | 1 | 1 | 0.9995 | 0.9999 | 0.9960 | 0.9624 | **0.9081** | 0.8453 |
| Graph Value 8 | 0.9999 | 1 | 1 | 0.9993 | 0.9978 | 0.9920 | 0.9639 | **0.9056** | 0.8471 |
| Graph Value 9 | 0.9998 | 0.9995 | 0.9998 | 0.9997 | 0.9981 | 0.9867 | 0.9882 | **0.9015** | 0.8385 |

**Table 5** Normalized mutual information (*NMI*) Benchmark model case 4.

| NMI | Relations Value 1 | Relations Value 2 | Relations Value 3 | Relations Value 4 | Relations Value 5 | Relations Value 6 | Relations Value 7 | Relations Value 8 | Relations Value 9 |
|---|---|---|---|---|---|---|---|---|---|
| **Graph Value 1** | 0.9996 | 0.9973 | 0.9927 | **0.9897** | **0.9808** | 0.9839 | 0.9764 | 0.9794 | 0.9748 |
| **Graph Value 2** | 0.9999 | 0.9996 | 0.9931 | 0.9922 | **0.9857** | 0.9948 | **0.9860** | 0.9728 | 0.9775 |
| **Graph Value 3** | 0.9999 | 0.9983 | 0.9955 | 0.9933 | 0.9989 | 0.9933 | **0.9899** | 0.9755 | 0.9786 |
| **Graph Value 4** | 0.9998 | 0.9981 | 0.9987 | 0.9933 | 0.9987 | 0.9969 | **0.9885** | 0.9808 | 0.9746 |
| **Graph Value 5** | 0.9997 | 0.9995 | 0.9946 | 0.9981 | 0.9965 | 0.9965 | 0.9969 | **0.9791** | 0.9765 |
| **Graph Value 6** | 0.9996 | 0.9999 | 0.9980 | 0.9984 | 0.9992 | 0.9943 | **0.9875** | 0.9741 | 0.9771 |
| **Graph Value 7** | 0.9999 | 0.9986 | 0.9988 | 0.9997 | 0.9982 | 0.9989 | 0.9917 | **0.9792** | 0.9754 |
| **Graph Value 8** | 0.9989 | 0.9990 | 0.9982 | 0.9992 | 0.9991 | 0.9986 | **0.9814** | 0.9772 | 0.9669 |
| **Graph Value 9** | 0.9997 | 0.9986 | 0.9985 | 0.9963 | 0.9977 | 0.9979 | **0.9845** | 0.9754 | 0.9716 |

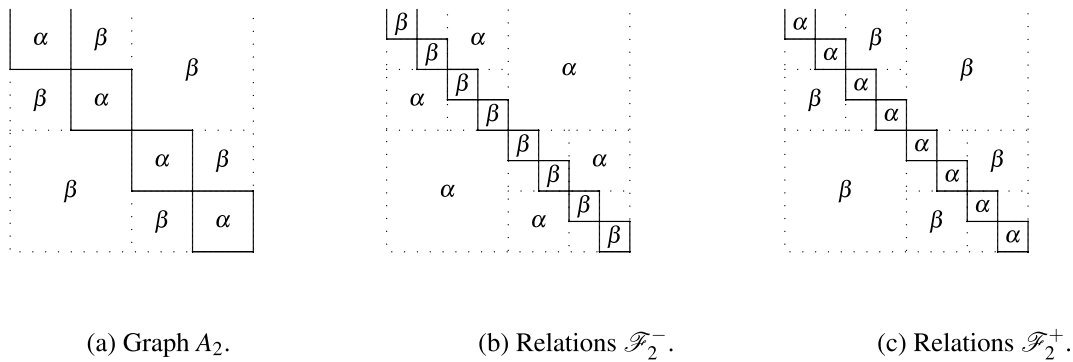

(a) Graph $A_2$.     (b) Relations $\mathscr{F}_2^-$.     (c) Relations $\mathscr{F}_2^+$.

**Figure 4** Benchmark model. Case 2.

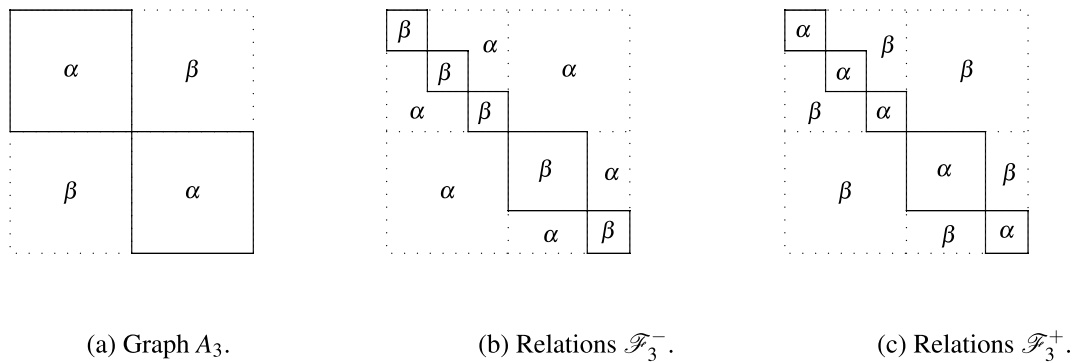

(a) Graph $A_3$.     (b) Relations $\mathscr{F}_3^-$.     (c) Relations $\mathscr{F}_3^+$.

**Figure 5** Benchmark model. Case 3.

well despite the complexity of the model. In fact, although we work with an asymmetric structure, the groups obtained resemble very well the gold partition, being affected just when the relations matrix becomes sparser.

### 5.2.4. Benchmark model. Case 4

Now we apply our algorithm in a very difficult model. Not only is the structure of the relations so asymmetric, but also the groups of the graph are somewhat small (see Figure 6). The adjacency matrix $A_4$ is built in the same way as it is built matrix $A_2$ in Subsection 5.2.2. Regarding relations matrices $\mathcal{F}_4^-$ and $\mathcal{F}_4^+$, both are built with 8 embedded clusters, whose expected sizes are $|C_1| = 40$, $|C_2| = 24$, $|C_3| = 64$, $|C_4| = 21$, $|C_5| = 22$, $|C_6| = 21$, $|C_7| = 32$, and $|C_8| = 32$. Despite this complex structure, our algorithm is able to recover the information embedded in the standard model successfully, as you can see in Table 5. Once again, the scarcity of the relations matrix is the point that most affects the obtained result.

## 6. CONCLUSIONS

Given a set of elements between which there are some relationships, one of the most used mathematical tools for its representation are networks. The inconvenient of networks is that we cannot model all the information involved in a real-life problem with them. Here lies the importance of working with graphs which can include some additional information. This approach was first managed in [12], where it was proposed the idea of considering affinity relations among the nodes by means of extended fuzzy graphs. Nevertheless, not only positive evidences appear when modeling a real-life problem, but also negative evidence. In order to deal with this type of situation, in [7] it was introduced the idea of working with extended bipolar fuzzy graphs, $\tilde{G} = (V, E, \mu^b)$.

In a broader context, the key point of this paper is related to community detection problems based on multiple bipolar fuzzy measures. It is, in the resolution of clustering problems in which there exists some additional information about the relation among the nodes, which is defined by multiple bipolar fuzzy measures, and which is absolutely independent of the structure of the graph.

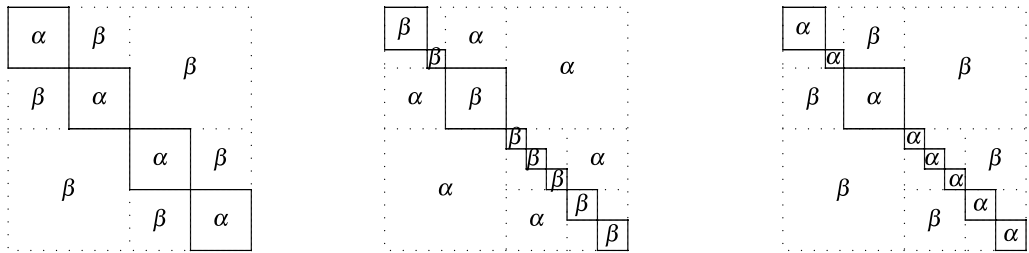

**Figure 6** | Benchmark model. Case 4.

We start this paper with the definition of several concepts: the extended bipolar fuzzy graph and the bipolar weighted multi-graph associated with a bipolar fuzzy measure. Then, we generalize the notion of extended bipolar fuzzy graph, and the bipolar weighted multi-graph associated with a bipolar fuzzy measure, to the field of having multiple bipolar fuzzy measures. So, we introduce two notions: extended multiple bipolar fuzzy graph and the bipolar weighted multi-graph associated with multiple bipolar fuzzy measures. Then, focusing on OWA operators [34–37], we propose several "group" notions, based on multiple bipolar fuzzy relations. Both elements will be crucial for dealing with community detection problems in which there is some additional information about the relation between the nodes given by multiple bipolar fuzzy measures.

Apart of this, in Section 4 we provide a detailed explanation of the performance of Louvain algorithm. Then we propose a modification of this method, the Duo Louvain Algorithm, based on Louvain's one. The new algorithm can differentiate two types of information when solving community detection problems (one used to find "possible" clusters and another to calculate the modularity optimization). To illustrate it, we provide a pseudocode (see Algorithm 2). We propose a particular application of Duo Louvain Algorithm for solving community detection problems with some additional information given by multiple bipolar fuzzy measures. We suggest the resolution of this type of problems, which are modeled by means of an extended multiple bipolar fuzzy graph $\tilde{G} = (V, E, (\mu^{b_1}, \ldots, \mu^{b_s}))$ with the *Multiple Bipolar Duo Louvain* Algorithm (see Algorithm 3). Let us remark that, as far as we know, these type of problems have never been faced, so the philosophy we follow in this paper is absolutely new.

In order to quantify how each node of the graph is affected by the presence/absence of any other node in its clusters, depending on their relation in $(\mu^{b_1}, \ldots, \mu^{b_s})$, we propose the consideration of the Shapley value [27]. Hence, we aggregate the fuzzy measures $\mu^{-\ell}$ and $\mu^{+\ell}$ into the matrices $\mathcal{F}^{-\ell}$ and $\mathcal{F}^{+\ell}$ respectively, which are used in Algorithm 3.

Other important point of this article is the development of a wide analysis of the performance of our algorithm. To carry one with it, we provide some computational results with which we test the "goodness" of the Algorithm 3 (see Section 5 for more details). These tests consist in the generation of four different benchmark models [4,16,40], in combination with the calculation of the *NMI* [17]. With this analysis, we measure the ability of our algorithm to recover an embedded cluster structure for a synthetic graph.

Tables 2–5 allow us to affirm that our algorithm provides very good results.

Apart from this consideration of the contributions included in this paper, clearly devoted to community detection problems in graphs, let us emphasize the there are many more applications of this research. One of them is about the field of fuzzy relation equations. This type of fuzzy relation equations, are deeply analyzed in [43–45], comprises situations in which some unknown variables appear together with their logical negations, something closely connected with each pair of fuzzy measures $(\mu^{-\ell}, \mu^{+\ell})$.

As further work, we propose different applications of *Duo Louvain* Algorithm, and, in particular, of the *Multiple Bipolar Duo Louvain* Algorithm, considering different operators $\Phi^-$, $\Phi^+$, and $\psi$. We will check the consistency and the stability of the method here developed [46,47]. It is well established that the Louvain algorithm performs well with very large networks [48,49], so we will also work with real large databases. Specifically, we will consider a Twitter extraction to analyze how the different profiles are grouped in communities depending on their political opinion.

These two points are supposed to be our immediate further research. We would also like to extend our idea to FCMs [14]. FCM, usually used as decision support tools, are defined by nodes called concept and weighted edges called relationships. Then, an easy vision of FCM is considering them as directed and weighted networks whose edges defines cause and effect relations among the nodes, and the degree of these relations [13].

Moreover, we are working on the inclusion of additional information given by some fuzzy measure in the framework of different community detection algorithms, specifically, the method proposed by Girvan and Newman [30].

## ACKNOWLEDGMENTS

This research has been partially supported by the Government of Spain, Grant Plan Nacional de I+D+i, MTM2015-70550-P, PGC2018096509-B-I00, and TIN2015-66471-P.